\documentclass[aps,prl,showpacs,twocolumn,amsmath,amssymb]{revtex4-1}

\usepackage{graphicx}
\usepackage{dcolumn}
\usepackage{bm}

\preprint{APS/123-QED}

\begin{document}

\title{Spatio-temporal Fermionization of Strongly Interacting 1D Bosons}
\author{Vera Guarrera}
\email{guarrera@physik.uni-kl.de}
\author{Dominik Muth}
\author{Ralf Labouvie}
\author{Andreas Vogler}
\author{Giovanni Barontini}
\author{Michael Fleischhauer}
\author{Herwig Ott}
\affiliation{Research Center OPTIMAS and Fachbereich Physik, Technische Universit\"at Kaiserslautern, 67663 Kaiserslautern, Germany}

\date{\today}

\begin{abstract}
Building on the recent experimental achievements obtained with scanning electron microscopy on ultracold atoms, we study one-dimensional Bose gases in the crossover between the weakly (quasi-condensate) and the strongly interacting (Tonks-Girardeau) regime.
We measure the temporal two-particle correlation function and compare it with calculations performed using the Time Evolving Block Decimation algorithm. More pronounced antibunching is observed when entering the more strongly interacting regime. Even though this mimics the onset of a fermionic behavior, we highlight
that the exact and simple duality between 1D bosons and fermions does not hold when such dynamical response is probed. The onset of fermionization is also reflected in the density distribution, which we measure \emph{in situ} to extract the relevant parameters and to identify the different regimes. Our results show agreement between experiment and theory and give new insight into the dynamics of strongly correlated many-body systems.
\end{abstract}

\pacs{03.75.Hh, 03.75.Kk, 05.30.Jp}

\maketitle
Interactions between particles are of particular importance in one-dimensional (1D) systems as they lead to strong quantum correlations. For ultracold bosonic atoms, reduced dimensionality and control of interactions can be achieved experimentally by quantum optics tools. This has lead to the observation of the strong-interaction regime of bosonic atoms \cite{weiss2004,bloch2004}.  
One-dimensional ultracold gases offer moreover a unique test bench for theory, as they are among the few many-body systems which can be described on the basis of integrable hamiltonians such as the Lieb-Liniger (LL) model \cite{liebliniger,yangyang}. While the integrability provides exact benchmarks for a comparison with experiments, important quantum characteristics such as higher-order correlation functions remain a challenge \cite{revmodphys,kheruntsyan2008}. Their measurement would provide important information about the many-body system beyond that accessible from density profiles 
in coordinate or momentum space \cite{correlation}. Only local second and third order correlation functions, $g^{(2)}(0,0)$ and $g^{(3)}(0,0)$, have been experimentally investigated in 1D via indirect diagnostics \cite{weiss2005,haller2011,bouchoule2011}. 
Of even larger interest are {\it temporal} correlations as they probe the nature of excitations, which goes beyond the characterization of quantum properties of the ground or thermal state of the system. 
In order to directly access correlation functions \emph{in situ}, spatially resolved single atom sensitive detection methods are well suited \cite{guarrera,kuhr2011}.
In this work, we use scanning electron microscopy to study one-dimensional tubes of ultracold bosonic atoms in the crossover between the weakly and the strongly interacting regime. 
We characterize the 1D systems by performing \emph{in situ} measurements of the spatial density distribution, which allows for the determination of the relevant parameters and the identification of the different interaction regimes.
The complete \emph{temporal} two-particle correlation function
\begin{eqnarray}
&&g^{(2)}(\xi=x-x_0,\tau=t-t_0)= \nonumber \\
&&\frac{\langle \hat{\Psi}^\dagger(x_0, t_0) \hat{\Psi}^\dagger(x, t) \hat{\Psi}(x, t)\hat{\Psi}(x_0, t_0) \rangle}{\langle \hat{\Psi}^\dagger(x_0, t_0) \hat{\Psi}(x_0, t_0)\rangle\langle \hat{\Psi}^\dagger(x, t)\hat{\Psi}(x, t)\rangle}
 \label{eq:g2}
\end{eqnarray}
is then measured for $\xi=0$. Here $\widehat{\Psi}$ are the bosonic field operators and $\langle ... \rangle$ indicates the quantum mechanical average. The results are finally analysed on the basis of the LL model, solved by numerical methods.

The usual classification of the 1D regimes for trapped Bose gases, under conditions that justify the local density approximation (LDA), is based on the dimensionless interaction parameter $\gamma(x)$ at the trap center and on the temperature $T$ \cite{olshanii2001,kheruntsyan2005}. The space dependent LL parameter is defined as $\gamma(x)=mg/\hbar^2 \rho(x)$, with $m$ the mass of the particle, $\rho(x)$ the density and $g\simeq 2a\hbar\omega_r$ the 1D coupling constant for $\vert a \vert<a_r$, where $a$ is the three-dimensional $s$-wave scattering length, $\omega_r$ the frequency of the radial harmonic confinement and $a_r=\sqrt{\hbar/m\omega_r}$. For $\gamma(0)\ll1$ and $T$ below the degeneracy temperature $T_d=\hbar^2\rho^2(0)/2mk_B$ \cite{kheruntsyan2005}, with $k_B$ the Boltzmann constant, a weakly interacting quasi-condensate phase is predicted. The spatial density profile in a harmonic trap is expected to be well described by a Thomas-Fermi (TF) parabola \cite{olshanii2001} and $g^{(2)}(0,0)\approx1$ \cite{kheruntsyan2005}. In the opposite limit, $\gamma(0)\gg1$, the 1D gas is strongly interacting and approximates a Tonks-Girardeau (TG) gas. The density profile is a square root of a parabola and $g^{(2)}(0,0)\ll1$, which indicates an effective reduction in the overlap of the particle wavefunctions, resembling the fermionic exclusion principle. 
\begin{figure}[t!]
\begin{center}
\includegraphics[width=0.3\textwidth]{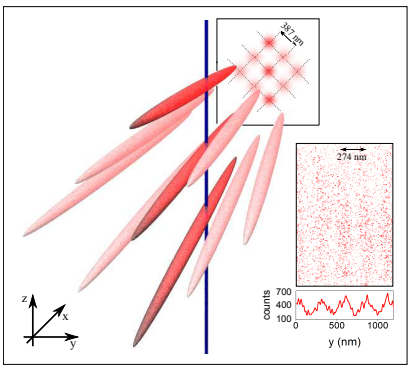}
\end{center}
\caption{(Color online) Schematic of the experiment. The 2D blue-detuned lattice realizes a system of parallel 1D tubes of ultracold bosons. When the electrons (blue line) collide with the atoms they lead to the production of ions which are subsequently detected. The inset shows a zoom on the sum of 100 images of the 1D tubes. The integrated density along the $x$ direction is also presented. In this picture the total atom number is $\sim 10^4$, the current of the electron beam is $20$ nA and its FWHM=$120(10)$ nm.} \label{fig:system}
\end{figure}

In our experiment, a Bose-Einstein condensate (BEC) of about $8\times10^4$ $^{87}$Rb atoms is created in the optical dipole trap realized by a focused CO$_2$ laser beam. Once the BEC is produced, its final atom number is accurately adjusted by scanning the electron beam on the outer part of the cloud. In this way we control the atom number and we selectively discard the warmer particles, effectively reducing the temperature of the sample (to less than $10$ nK). For the experiments reported here, we prepared two different sets of BEC samples with $10(2)\times10^3$ (10k) and $60(5) \times 10^3$ (60k) atoms respectively. In order to create the 1D atomic tubes, we adiabatically superimposed to the CO$_2$ dipole trap a two-dimensional (2D) blue-detuned optical lattice, realized by a pair of retroreflected laser beams with wavelength $774$ nm and waist $630$ $\mu$m (see Fig.~\ref{fig:system}). The final frequencies in the tubes for the two sets of measurements are $\omega_{a}/2\pi=8$ Hz, $\omega_{r}/2\pi=56$ kHz (10k atoms) and $\omega_{a}/2\pi=11$ Hz, $\omega_{r}/2\pi=40$ kHz (60k atoms), where $\omega_{a}$ is due to the axial confinement of the dipole trap. The number of tubes created varies from $\sim 100$ to $280$, with a maximum occupation number in the centre of about $170$ and $500$ atoms respectively. Changing the density of the initial three-dimensional system and the radial confinement $\omega_r$, we are thus able to experimentally access two different 1D regimes in the crossover between the strongly and the weakly interacting limit: $\gamma(0)\backsimeq2$ in the central tube for the set with 10k atoms and $\gamma(0)\backsimeq0.5$ for the set with 60k atoms.

In order to characterize in detail our experimental 1D systems, we image the entire cloud by scanning it with an electron beam of $6$ keV energy, $60$ nA current and $240(10)$ nm FWHM \cite{Gericke2008}. At the end of the scan the number of atoms remaining in the trap is probed by time-of-flight absorption imaging and the BEC is checked to be pure. For each set of measurements $\sim 300$ images of the cloud are summed. Each scanning line of the image along the $x$-direction corresponds to the sum of the density profiles of the 1D tubes displaced in a vertical row along the direction of the electron beam, see Fig.~\ref{fig:system}. 

\begin{figure*}[t!]
\includegraphics[width=0.48\textwidth]{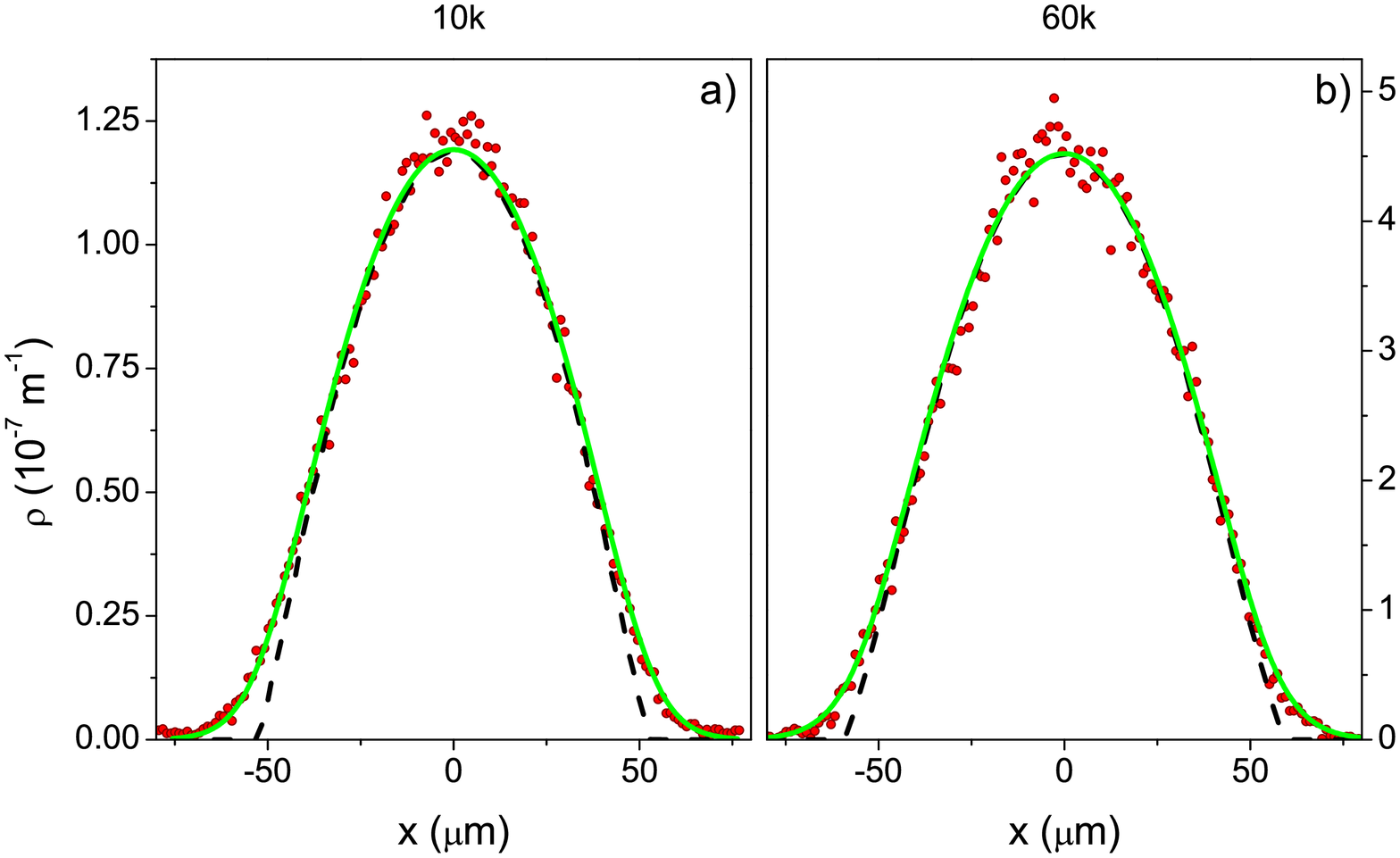}
\includegraphics[width=0.48\textwidth]{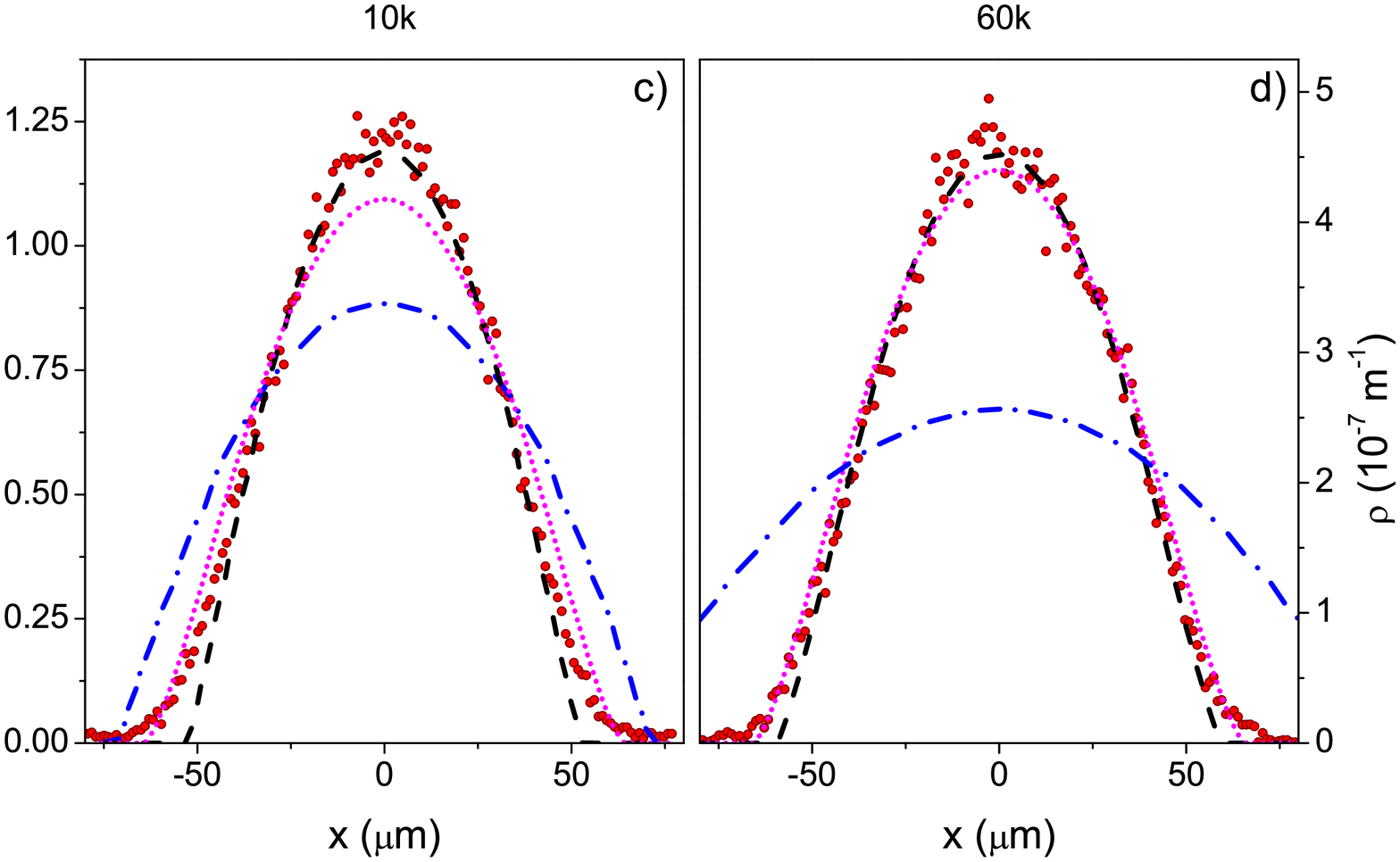}
\caption{(Color online) \emph{In situ} spatial density distribution summed over $8$ and $11$ tubes for the measurement sets with 10k atoms, (a) and (c), and 60k atoms, (b) and (d), respectively. Data (red dots) are the result of the sum of $\sim 300$ pictures. In (a) and (b) the green solid lines are fits using YY theory and the dashed black lines are $T=0$ LL profiles. In (c) and (d) the $T=0$ LL profiles are compared with those calculated with the TF (magenta dotted line) and with the TG theory (blue dashed-dotted line).} \label{fig:density}
\end{figure*}
For the two sets of measurements, the density profiles of the central row of the images, corresponding to the sum of respectively $\sim 8$ and $11$ tubes along the $z$-direction, are shown in Fig.~\ref{fig:density}. 
The results are first compared with the exact Yang-Yang (YY) solution, where the only free parameters are the total atom number and the temperature. We numerically solve the YY equations for a uniform 1D gas, for a given temperature, and we then apply the LDA to account for the axial confinement $V(x)=m\omega_a^2x^2/2$ \cite{kheruntsyan2005}. The resulting density profile is then obtained as a sum over the different tubes, where the number of tubes and atoms per tube are determined on the basis of the TF distribution of the initial BEC. The final curves show excellent agreement with the experimental data. Notably this allows for an accurate determination of $\gamma(0)$ for the different tubes and of the temperature of the system. For the first set of measurements (10k atoms) we find that $\gamma(0)$ ranges from $2.2(2)$ to $13(1)$ in the different tubes along the $z$-direction, resulting in the weighted average $\overline{\gamma(0)}=2.7(3)$. The fitting parameters are $N=9.3(0.7)\times10^3$ atoms and $T=9(1)$ nK. For the second set (60k atoms): $\gamma(0)=0.58(5)-2.3(2)$ and $\overline{\gamma(0)}=0.76(7)$ with $N=52(5)\times10^3$ and $T=22(2)$ nK. For further confirmation of the extracted temperatures, we additionally measure $T$ in two ways: with a gaussian fit of the density in the outermost tubes along the $y$-direction and by calculating the average degeneracy temperature at which the density profile of the outer tubes starts to match with a normal distribution, when moving along the $y$-axis. We find: $T=10(2)$ nK and $T=8(1)$ nK for 10k atoms, $T=27(6)$ nK and $T=28(5)$ nK for 60k atoms. In Fig.~\ref{fig:density}(a)-(b), we also show numerical solutions for the LL theory at $T=0$ with LDA, calculated for the same $\gamma(0)$ values extracted from the YY fit. The two models agree well, showing only a small discrepancy at the wings. This is due to the fact that the measured temperatures $T$ are below the weighted average of $T_d$ over the tubes, i.e. $T/\overline{T_d}=0.7(1)$ (10k atoms) and $T/\overline{T_d}=0.3(1)$ (60k atoms). To better clarify the regime of our measurements, in Fig.~\ref{fig:density}(c)-(d) we compare the profiles obtained with the LL model with those calculated by solving the Gross-Pitaevskii equation in the TF approximation and by solving the TG model, with the same atom number \cite{olshanii2001}. For the lower $\gamma$, Fig.~\ref{fig:density}(d), the actual density profile is very close to the TF, while the TG distribution is significantly distant: the system lies in the weakly interacting TF limit. Increasing $\gamma$ above $1$, Fig.~\ref{fig:density}(c), the experimental data and the exact 1D theory start to significantly deviate from the TF distribution. The TG profile, conversely, becomes closer: an intermediate regime towards the TG limit and the complete fermionization has been entered.

\begin{figure}[t!]
\begin{center}
\includegraphics[width=0.5\textwidth]{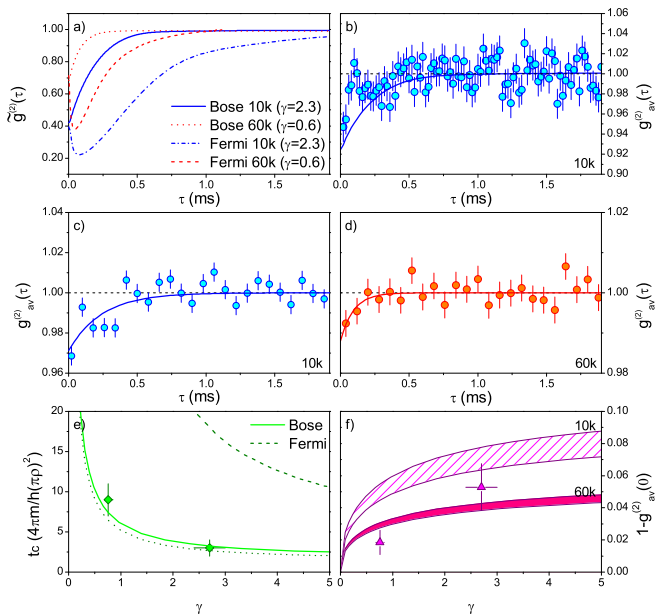}
\end{center}
\caption{(Color online) (a): $g^{(2)}(\tau)$ calculated using the TEBD algorithm together with the respective Fermionic counterpart for the single central tube. (b): Measured $g^{(2)}_{\rm av}(\tau)$ for 10k atoms. The solid line is obtained from TEBD numerical calculations of the independent tubes \cite{average}. The theoretical curve has no free parameters, since they are determined from the analysis of the density profiles. (c)-(d): Measured $g^{(2)}_{\rm av}(\tau)$ with binning time of $80$ $\mu$s for 10k and 60k. The solid lines are exponential fits to the data. (e): Correlation time $t_c$ as a function of $\gamma$. The dotted line are bare TEBD calculations, the solid line includes averaging over the electron beam size and the dashed line shows the fermionic theory. (f): Amplitude as a function of $\gamma$. The uniformly filled and patterned areas are the theoretical calculations according to \cite{average}. The extension of the area is due to an uncertainty $\pm1$ on the number of tubes.} \label{fig:correlation}
\end{figure}

When we proceed to the measurement of $g^{(2)}(\tau)$, the differences between the two experimental situations outlined above become even more evident. The observation of interaction induced antibunching in the two-particle correlation function is the most direct indication of a fermionic behaviour emerging in an ensemble of bosonic particles. The experimental procedure is similar to the one described in our earlier work \cite{guarrera}. Once the 1D systems are prepared, we investigate them by focussing the electron beam at the center of the cloud. What we measure is the average correlation function of two ionization events $g^{(2)}_{\rm av}(\tau)$ obtained by probing simultaneously $8$ or $11$ different tubes. For these measurements we set the electron beam current to $20$ nA and FWHM$=120(10)$ nm. The spatial resolution, set by the FWHM of the electron beam, is smaller than the minimum estimated spatial correlation length for the probed systems, i.e. $440$ nm (10k atoms) and $170$ nm (60k atoms). The $g^{(2)}_{\rm av}(\tau)$ is measured on the ion signal collected in $50$ ms. This is the longest holding time at which the system is still a pure BEC. We compute the correlation functions over about $12000$ (10k atoms) and $2000$ (60k atoms) repetitions of the experiment. In Fig.~\ref{fig:correlation}(b) we report the details of the measured $g^{(2)}_{\rm av}(\tau)$ for the set with 10k atoms, where antibunching is visible with an amplitude $1-g^{(2)}_{\rm av}(\tau=0)=0.05(1)$. For $\tau>0$ the shape of the signal suggests the existence of two timescales on which the correlations relax: a short one ($\sim 100$ $\mu$s) and a longer one ($\sim 500$ $\mu$s). We first discuss the longer timescale and, to this purpose, we show the data with 80 $\mu$s bin time in Fig.~\ref{fig:correlation}(c)-(d). From these data, using an exponential fitting, we extract the correlation times $t_c=240(80)$ $\mu$s (10k) and $100(40)$ $\mu$s (60k), defined as the times corresponding to the $1/e$ reduction of $1-g^{(2)}_{\rm av}(\tau)$. 

The $g^{(2)}(\xi,\tau)$ function in the ground state can be theoretically calculated for each single tube, by simulating the dynamics of the system using exact numerical methods. We note that the use of a $T=0$ theory is justified in this case since the majority of the investigated tubes are below $T_d$ \cite{tzero}. In order to perform the simulations, we discretize the LL model \cite{Muth2010}. This lattice problem can then be treated using the Time Evolving Block Decimation (TEBD) algorithm \cite{Vidal2003}, which is one of the extensions of the Density Matrix Renormalization Group to time evolution \cite{Daley2004} and is based on a Matrix Product State (MPS) ansatz \cite{Schollwock2011}. We perform the calculation on a gas of $N=25$ particles confined in a harmonic potential. $g^{(2)}(\xi,\tau)$ is evaluated in the center of the tube at $x_0=0$, according to the experiment. 
The numerical procedure consists of first calculating the interacting ground state $\vert0\rangle$ at fixed particle number $N$ \cite{Muth2010}. We then apply the annihilator at $x_0=0$, which is straightforward in the discretized MPS representation, resulting in a $N-1$ particle state $\vert \Psi_0 \rangle =  \hat{\Psi}\vert0\rangle$. This state is finally propagated in time. 
The spatio-temporal correlation function $g^{(2)}(\xi,\tau) = \rho^{-2}\ \langle \Psi_{\tau} \vert \hat{\rho}(x)\vert \Psi_{\tau} \rangle$ is then the expectation value of the local density operator $\hat{\rho}(x) = \hat{\Psi}^\dagger(x)\hat{\Psi}(x)$ in the time-evolved state $\vert \Psi_{\tau}\rangle=\exp(-i\hat{H}\tau/\hbar)\vert \Psi_{0}\rangle$, where $\hat{H}$ is the Hamiltonian of the system. The finite size of the system turns out to have no influence because all the excitations, travelling with the speed of sound, reach the edges of the cloud only at a time when the antibunching structure in $g^{(2)}(\tau)$ has already decayed.   
In addition we take into account the finite width of the electron beam by averaging over its Gaussian profile $W(x')$: $\tilde{g}^{(2)}(\tau) = \int_{-\infty}^\infty{\rm d}x'\ W(x')\ g^{(2)}(x', \tau)$, see Fig.~\ref{fig:correlation}(a). As the experiment measures several independent tubes at the same time, we calculate $g^{(2)}_{\rm av}(\tau)$ by averaging over the $\tilde{g}^{(2)}(\tau)$ corresponding to each tube \cite{average}. The uncorrelated signal coming from different independent tubes is responsible for the reduction of the antibunching amplitude from the value of $1-\tilde{g}^{(2)}(0)\simeq0.60$ ($0.40$) for the single tube to the global calculated $1-g^{(2)}_{\rm av}(0)\simeq0.07$ ($0.03$) in the systems with 10k (60k) atoms. Concerning the longer timescale, the agreement of the averaged theoretical curves with the experimental data in Fig.~\ref{fig:correlation}(b) is fairly good. For a quantitative comparison, we show in Fig.~\ref{fig:correlation}(e) and (f) the experimentally derived correlation times and amplitudes together with their theoretical counterparts. 
The correlation time $t_c$ rescaled by $1/\rho^2$ is independent on $\rho$ and is strongly depending on $\gamma$ in the regime of our measurements: increasing the interactions above $\gamma=1$ we observe a significant decrease of $t_c$. This behaviour is in good agreement with our experimental findings. The interpretation of $1-g^{(2)}_{\rm av}(0)$ as a function of $\gamma$  in the regime that we consider is mainly complicated by the average over a different number of tubes. This effect is indeed contributing to the increase of the amplitude for the higher value of $\gamma$. Anyway the comparison between the experimental and theoretical amplitudes is compatible within the error bars. We note that finite temperature effects cannot have a significant role in the reduction of the amplitude \cite{tzero}.

It is well known \cite{girardeau,gaudin,cheon} that 1D bosons can be mapped to interacting fermions for arbitrary interaction strengths. As a consequence the density distribution and density-density correlations of bosons are identical to that of the corresponding fermions. Conversely, this simple Bose-Fermi duality breaks down when the temporal pair correlation function is probed: the theoretical curves in Fig.~\ref{fig:correlation}(a) and (e) show indeed a notable difference between the fermionic and the bosonic case, which is confirmed by the experimental data.
The observable measured and calculated in this work, i.e. $g^{(2)}(\xi,\tau)$, is fundamentally different from the dynamical density-density correlations previously derived in Ref.~\cite{caux2006} because the bosonic field operators do not commute at different times. This is also the reason for Bose-Fermi duality to fail. 
We finally discuss the shorter timescale visible in Fig.~\ref{fig:correlation}(b), which corresponds to an additional modulation of the signal with frequency $\sim 5$ kHz. The energy scale introduced by the electron beam can be roughly estimated to be $\hbar^2/2m$FWHM$^2\simeq h\times4$ kHz, suggesting it to be the origin of such an oscillation. We note that a similar modulation is not visible, within the error bars, in the weakly interacting case and was never observed in a 3-dimensional degenerate sample \cite{guarrera}. This suggests this behaviour could be associated to the presence of strong correlations between the particles when $\gamma>1$.       

In summary, we have performed \emph{in situ} high-resolution measurements of the density profiles of few 1D tubes in the crossover between the weakly ($\overline{\gamma}\backsimeq0.76$) and the more strongly ($\overline{\gamma}\backsimeq2.7$) interacting regimes. We find excellent agreement in the comparison with a model based on the YY solutions and LDA, allowing us to extract the $\gamma(0)$ values of the tubes, the global atom number and the temperature. Comparison with the $T=0$ LL, TF and TG models allows us to clearly define the regimes of the measurements and to circumscribe the role of temperature.
We measured the temporal two-particle correlation function, observing genuine antibunching in correspondence to the entering of the more strongly interacting regime, characterized by evident fermionization of the bosonic particles already at intermediate values of $\gamma$. 
We have numerically calculated $g^{(2)}(\xi,\tau)$ and we find fairly good agreement with the experimental data, showing the exact duality following the Bose-Fermi mapping does not hold for this observable. This work paves the way to the study of complex dynamics as those resulting from quenches \cite{bernier2011} and to the investigation of thermalization processes in 1D \cite{rigol2009}.

\begin{acknowledgments}
We thank B. Schmidt for fruitful discussions and P. W\"urtz and F. Stubenrauch for technical support and experimental assistance.
We acknowledge financial support by the DFG within the SFB/TRR 49 and GRK 792. V. G. and G. B. are supported by Marie Curie Intra-European Fellowships. D. M. and R. L. acknowledge support by the MAINZ graduate school.
\end{acknowledgments}


\begin{references}
\bibitem{average}
The average is performed summing the signals, i.e. the ions, coming from the same tubes with the uncorrelated ones coming from the coupling of the different tubes: $g^{(2)}_{\rm av}(\tau)=\sum_{i=1}(N_i/N)^2 (\tilde{g}^{(2)}_i(\tau)-1)+1$, where $N_i$ is the atom number in the $i$-th tube, $N$ is the total number of atoms and $\tilde{g}^{(2)}_i(\tau)$ is the $\tilde{g}^{(2)}(\tau)$ calculate for the $i$-th tube with the $\gamma_i(0)$ values derived from the analysis of the density profiles. The pointing instability of the electron beam is also taken into account.
\end{references}
\end{document}